\documentstyle[aps,epsfig,prl,floats]{revtex}
\begin{document}
\draft
\twocolumn[\hsize\textwidth\columnwidth\hsize\csname@twocolumnfalse%
\endcsname
\title{Crossover and scaling in a two-dimensional field-tuned superconductor}
\author{Sumanta Tewari}
\address{Physics Department, University of California, Los Angeles, CA
90095--1547}
\date{\today}
\maketitle
\begin{abstract}
Using an analysis similar to that of Imry and Wortis, it is shown that the apparent first order superconductor to metal transition, which has
been claimed to exist at low values of the magnetic field in a two-dimensional field-tuned system at zero temperature,
can be consistently
interpreted as a sharp crossover from a strong superconductor to an inhomogeneous state, which is a weak superconductor. The true zero-temperature 
superconductor to insulator transition within the
inhomogenous state is conjectured to be that of randomly diluted $XY$ model. An explaination of the observed finite temperature
approximate scaling of resistivity close to the critical point is speculated within this model.  
\end{abstract}

]
\narrowtext

Two-dimensional (2D) zero-temperature magnetic field($H$)-tuned superconductor-insulator transition (SIT) 
has recently attracted a lot of attention\cite{yazdani,ephron,shimsoni,mason1,kapitulnik,mason2,mason3}. 
It is common to think of this transition as a prototypical continuous quantum phase transition (QPT) at
a critical magnetic field $H=H_c$ with a diverging correlation length $\xi \sim |H_{c}-H|^{-\nu}$, 
and associated scaling\cite{fisher1}. Here, $\nu$ is the correlation length exponent. The state on the
$H<H_c$ side of the transition is a superconductor where the vortices are localized in a so-called
vortex glass state, and the Cooper pairs are delocalized. The state on the other side of the transition
is an insulator, where the vortices form a superfluid (they follow Bose statistics as well 
in a dual representation\cite{dhlee1,dhlee2}), and the Cooper pairs are localized. Exactly $\it{at}$
the transition, both the vortices and the Cooper pairs are delocalized, and the system is supposed to 
show metallic behavior with a universal resistivity\cite{fisher1,fisher2,wen}.

  However, recent experiments on amorphous Mo-Ge films do not
  show scaling in the limit of very low temperatures $T\leq 100 mK$\cite{mason1}, do not show universal resistivity at the
  putative transition point, and, in fact, show signatures of zero temperature saturation of Ohmic
  resistance in a wide range of $H$ on both sides of the putative $H_c$\cite{yazdani,mason1}. The system thus appears to have a metallic ground
  state in almost all cases $\it{except}$ below a very low value of the magnetic field, $H_{\text {sm}}$ (`sm' for superconductor-metal), where true
  superconductivity is apparently recovered\cite{mason2}. Hysteresis is observed in the vicinity of $H_{\text {sm}}$. Phenomenologically,
  there seems to be a zero-temperature first order field-tuned superconductor to metal transition at
  $H\sim H_{\text {sm}}$. The conventional picture of the field-tuned SIT, then, needs to be seriously revisited. There
  is also experimental evidence of a metallic-like phase in thickness or disorder-tuned superconductor-insulator systems\cite{jaeger,chervenak}.
  The possibility of a metallic phase intervening between the insulating and the superconducting phases has been
  a subject of many recent theoretical papers\cite{shimsoni,denis,larkin,spivak}. In the following, we will discuss only the field-tuned systems;
  whether similar considerations can be applied to other systems as well is left for future study.
  
   We will first
  focus on the low field superconductor to metal transition in the field-tuned system and argue that it is a remnant effect of the effects of quenched disorder on
  the first order vortex lattice melting transition of the pure system at a higher value of $H=H_{\text{cr}}$.
  Effects of quenched disorder on first order transitions have a long history in the literature. Imry and Ma \cite{ma} showed
  that for disorder coupling as random fields conjugate to the order parameter, first order transitions are completely eliminated below a critical 
  dimension which depends on the symmetry of the order parameter. Imry and Wortis \cite{imry} then argued that for disorder coupling as 
  random bond (or random $T_c$ in a classical system), the discontinuities associated with a first order transition are eliminated. They also
  proposed a mechanism of how that happens and 
  indicated the possibility that for this kind of disorder-coupling the first order transition may reduce to a continuous one. Later, Aizenman
  and Wehr \cite{wehr} proved mathematical theorems conclusively proving these heuristic assertions. In this paper, we will closely follow
  the argument of \cite{imry}, but make the most use of a feature, often overlooked, that the elimination of discontinuities
  of a first order transition due to random-bond coupling of the disorder {\it{can}} start as a sharp crossover at $D=2$. Beyond  the sharp crossover and within the resulting inhomogenous
  state, the {\it{true}} thermodynamic phase transition can be percolative, which is a continuous transition and thus consistent with the established
  theorems.
  
  We will work in terms of field-induced vortices instead of Cooper pairs. The putative SIT is then the melting
  transition of the vortex solid to a vortex liquid, which, in fact, is a vortex superfluid\cite{fisher1}. In the limit of a clean
  system free of any impurities, this is the first order Abrikosov lattice melting transition at $n=n_{\text {cr}}$. 
  For an average impurity concentration $p$,
  let's assume that $n_{\text {cr}}$ is renormalized to $n_{\text {cr}}(p)$. At this level, $n_{\text {cr}}(p)$ is 
  an assumed first order transition point between
  a vortex solid phase and a vortex liquid phase. The vortex solid phase may be a vortex glass (see, however,\cite{dfisher}), or a phase where the
  vortices are trapped individually by the impurity potentials, or even some (as yet) unknown state where the vortices are localized
  (note that in the presence of random pinning by impurities, which couple to a crystalline solid as random fields, the translational long range
  order of the crystal is destroyed below $D=4$\cite{ma}).
  Our second assumption is that, in the vortex solid phase, ${\it local}$ fluctuations in the impurity concentration 
  modulate the ${\it local}$ $n_{\text {cr}}$. This is sensible (although we cannot prove it here), since the impurities, 
  along with the repulsive interaction among the vortices, are
  responsible for vortex-localization. This can also be seen as following
  from the dependence of $n_{\text {cr}}(p)$ on $p$. Local $p$ may fluctuate from the average $p$, and locally modulate the critical 
  concentration. 
  Later, we will argue, following\cite{imry}, that $n_{\text {cr}}(p)$ will be rounded in a real system
  by domain formation on both sides of the transition, and the hypothetical vortex solid will loose its meaning.
  This is the reason why a precise characterization of the vortex solid is irrelevant for the present purpose.
  This line of argument will, however, bring out the importance of a sharp crossover-concentration, $n_{\text {sm}}$, which we will identify with
  $H_{\text {sm}}$, beyond which the domains proliferate and the system shows appreciable low-temperature resistance\cite{shimsoni}. 
  The true macroscopic superconductor-insulator 
  transition at $H_c$ can then only be defined in terms of percolation, by that of randomly diluted Josephson Junctions between the domains
  of the vortex solid phases. Since the zero-temperature SIT in this framework is essentially geometric, and not a prototypical
  QPT, scaling at finite temperature close to the critical point is {\it not} immediate.
  We will speculate on the consequences of such a transition to finite temperature scaling of Ohmic resistivity.

  In order to clarify the said remnant effect, let us flesh out the arguments for the effects of disorder on a first order
  transition.
  In our formulation in terms of a hypothetical vortex solid (and not a crystalline lattice), the analysis of \cite{imry}
  directly applies. It will pay to briefly go through the argument, however, since we will use the final
  result to explain the experiments. It will also help develop the alternative framework of SIT presented here (see also \cite{shimsoni}).
  Since the correlation length $\xi_{{\text cr}}$ remains finite at the transition,
  only regions of finite area $\sim \xi_{{\text cr}}^2$ are correlated close to the transition. The ${\it typical}$ density fluctuation of
  the impurities in a correlated region of linear dimension $\xi_{{\text cr}}$ is \cite{imry}
  \begin{equation}
  \Delta p \sim [p(1-p)]^{\frac{1}{2}}\xi_{{\text cr}}^{-1}.
  \end{equation} 
  Density fluctuations of other
  magnitudes are possible, but their probabilities are exponentially suppressed\cite{jeffreys}.
  So a correlated region has impurity concentration $p+\Delta p \gtrsim p \gtrsim p-\Delta p$ with maximum probability.
  Now suppose we are approaching the transition
  point by increasing the perpendicular magnetic field $H$, thus increasing $n$. As long as $n<n_{\text {cr}}(p)$ the system is 
  nominally in the solid phase. But a typical correlated region may already be
  in the $\it{wrong}$ phase due to fluctuations in the impurity concentration. The $\it{wrong}$ phase is a vortex liquid, which is not yet a superfluid 
  since the size is finite. Thus, if there were no interfacial
  surface energy between the solid and the liquid phases, then the first order transition would have had a smearing width
  
   \begin{equation}
\Delta n_{\text {cr}}=|\frac{dn_{\text {cr}}(p)}{dp}|\Delta p,
\end{equation}
   and the domain-sizes would have been of the order of $\xi_{{\text cr}}$. 
   
  But if we take into account the interfacial surface energies, as we must, then the domain-sizes can be much bigger than $\xi_{{\text cr}}$. This is
  possible, because the system has to balance the loss in the surface energy by gaining sufficient volume energy. By creating
  domains comprising several correlated regions with linear dimensions $\xi_{{\text cr}}$, it can maximize the gain in volume energy. One can show
  that\cite{imry},
  as one approaches the critical concentration $n_{\text {cr}}(p)$ by varying $H$, clusters of a general size $l$ will be first unstable
  at a concentration $n_{\text {cr}}(p)-\Delta n$ whenever
  
  \begin{equation}
  \Delta n \leq g(l),
   \label{condition}
\end{equation}
   where $g(l)$ is a function of the domain size $l$,
   
   \begin{equation}
   g(l)=[p(1-p)]^{\frac{1}{2}}l^{-1}|\frac{dn_{\text {cr}}(p)}{dp}| - \frac{C\sigma n_{\text {cr}}(p)l^{-2}}{f(p)}.
   \label{imp}
\end{equation}
  
\begin{figure}[bht] 
   \narrowtext 
   \begin{center} 
   \leavevmode 
   \noindent 
   \hspace{0.3 in} 
   \centerline{\epsfxsize=2.7in \epsffile{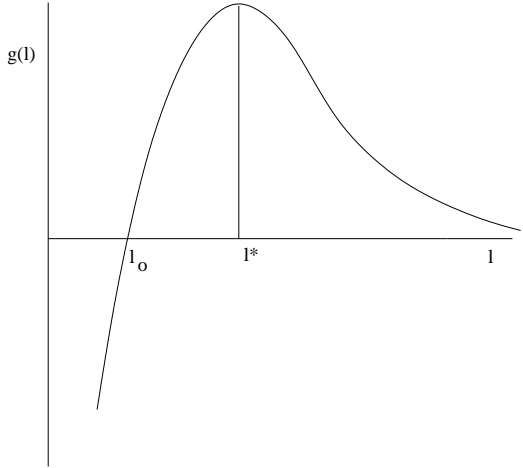}} 
   \end{center} 
   \caption 
{A schematic sketch of the function $g(l)$ for $D=2$. 
 $l=l^*$ is a point of inflection, which is a maximum.} 
   \label{fig1} 
   \end{figure}

  Here $\sigma$ is the interfacial surface tension, $C$ is a geometrical factor of order one, and $f(p)$ is a function of
  $p$, $f(p)=[n\frac{\partial (f_{l}-f_{s})}{\partial n}]_{p, n_{\text {cr}}(p)}$, where $f_{l}$ and $f_{s}$ are the free energy densities
  of the vortex liquid and solid phases respectively. 
  The formula in Eq.~\ref{imp} is for $D=2$.
  The first term is a measure of the volume energy gain by creating the liquid phase in a domain of size $l$, and the 
  second term comes from the surafce energy lost by doing it. Equation \ref{condition} is the first order analog of the Harris
  criterion\cite{harris}.
   Both the rounding width $\Delta n$ and the domain-size
  $l^{*}$ are now determined by the single equation,
  
  \begin{equation}
  \Delta n=g(l^{*}),
  \label{cond}
\end{equation}
   where $l^{*}$ maximizes $g(l)$. 
   This does not mean, however, that for $n<n_{\text {cr}}(p)-g(l^*)$ the system is free of domains. 
   What Eq.~\ref{cond} simply implies is that within the rounding width
   $\Delta n=g(l^{*})$ of $n_{\text {cr}}(p)$, the ${\it typical}$ domains of the vortex liquid of size $l^*$ will proliferate in the solid. This is
   because here the system is nearer to the transition point of the clean system.
   For the dimension of interest in the present paper, $D=2$, 
   we give in Fig.~\ref{fig1} a schematic
   sketch of $g(l)$.
   
    The basic point of observation, which is also one of the central points in the present paper,
    is that $g(l)$ rises from zero at $l=l_0$ and goes back to zero for $l\rightarrow \infty$, and so is ${\it peaked}$ at the point of inflection  
    $l=l^{*}$. 
     Consequently, as we increase the density
   of the vortices, $n$, by increasing $H$, domains of vortex liquid of size $l^{*}$ can first appear sharply (sharpness depends on the width of
   the peak of $g(l)$ at $l^{*}$, which in turn depends on the parameters of the system such as $\sigma$ and $f(p)$)
   at $n=n_{\text {sm}}=n_{\text {cr}}(p)-\Delta n=n_{\text {cr}}(p)-g(l^{*})$.
    Also, if $\xi_{{\text cr}}$, the finite correlation
   length of the would-be first order transition, is smaller than $l^{*}$, then the domain-size at the sharp crossover
   is not bounded by $\xi_{{\text cr}}$, rather it is $l^{*}$, which can be much bigger than $\xi_{{\text cr}}$ again depending on the system parameters.
   As $n$ is increased more, the function $g(l)$ also satisfies Eq.~\ref{condition} for values of $l$ other than $l^{*}$, and, in particular,
   the domain sizes can be arbitrarily large as $\Delta n \rightarrow 0$. The typical domain size, however, will always be
   $l^{*}$, since by creating domains of that size the system gains maximum energy.
   
   
   A domain of the vortex solid phase is $\it{almost}$ a superconductor, since the vortices are immobile. Similarly, a domain of 
   the vortex liquid phase is $\it{almost}$ an insulator. Low-temperature transport properties of a phase with inhomogenous admixture of such domains have been
   recently studied in\cite{shimsoni}. Because of quantum tunneling of the vortices among the domains of vortex liquid, 
   finite resistance arises in the system. The formula given in\cite{shimsoni} for this residual saturation
   tunneling resistance produces good fits to the experimental data \cite{ephron,mason1}. One of our aims in this paper has been to clarify the origin
   of such an inhomogenous state and to show that it can first occur sharply with increasing $H$. For values of $H$ less than the
   crossover value, $H_{\text {sm}}$, the system is a strong superconductor with no appreciable residual resistance.
   This is because the domains of the vortex solid are large (the domains of the vortex liquid are rare), 
   which makes the tunneling of the vortices impossible. 
     For $H \gtrsim H_{\text {sm}}$ the system has
   domains of typical size $l^{*}$ and the resulting weak superconductor has low-temperature tunneling resistance. Because of the proliferation of the 
   domains, hysteresis
   may be observed in the vicinity of $H_{\text {sm}}$, and the sharp crossover may $\it{look}$ like a first order superconductor 
   to metal quantum phase transition. This is the remnant effect alluded to before.
   
   Experiments of\cite{yazdani,ephron,mason1} have observed approximate scaling of resistivity inside the metallic-like
   phase in the phase diagram. This is indicative of a continuous phase transition at some critical value of $H=H_c > H_{\text {sm}}$.
   This transition, however, must be defined in the sense of percolation, since the vortices are not in the perfect solid phase.  
    As we have seen, for $H \gtrsim H_{\text {sm}}$ the sample is in an inhomogenous admixture of domains of $\it {almost}$
   superconductors and $\it {almost}$ insulators of typical size $l^{*}$. As $H$ increases towards $H_{\text{cr}}$ the insulating domains proliferate
   since the energy gain by creating them increases. The requisite continuous phase transition, then, can be that of
   two-dimensional randomly diluted Josephson Junctions\cite{granato}. The fact that the $I-V$ curves are nonlinear at relatively low
   currents attests to the presence of Josephson Junctions in this region of the phase-diagram\cite{mason2}.
   Ignoring capacitive charging energies of the individual domains, which is reasonable for large enough domain-sizes,
   this system is isomorphic to randomly diluted $XY$ model in two dimensions\cite{lub}.
   
\begin{figure}[bht] 
   \narrowtext 
   \begin{center} 
   \leavevmode 
   \noindent 
   \hspace{0.3 in} 
   \centerline{\epsfxsize=2.7in \epsffile{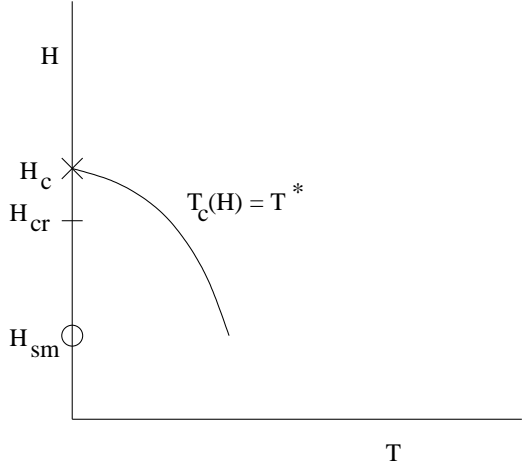}} 
   \end{center} 
   \caption 
{A schematic phase diagram of the system within the diluted $XY$ model. $H_c$ is the ${\it true}$ critical point, close to which scaling 
should apply. $H_{\text{cr}}$ is the avoided first order melting point. The sharp crossover happens at $H_{\text {sm}}$, beyond which
domains of size $l^*$ proliferate. $T_c(H)=T^*$ is the transition temperature of the diluted $XY$-model, but only a crossover temperature
for the real two-dimensional system. Below this temperature, in
a so-called renormalized-classical phase, the properties are dominated by dilute thermal excitations above the superconducting ground state.}
   \label{fig2} 
   \end{figure}

   These models exhibit classical phase transitions.
   When the sites or the bonds are randomly diluted, the transition temperature 
   $T_{c}(x)$ decreases monotonically
   with the concentration $x$ of occupied sites (or bonds). At the percolation threshold $x=x_c$, which corresponds to $H=H_c$ in the present
   problem, an infinite connected cluster of occupied sites ceases to exist, and $T_{c}(x)=0$. The point $P: x=x_c \hspace{1.0 mm}(H=H_{c}), T=0$ is a multicritical point
   where long range thermodynamic and geometric order develop simultaneously\cite{stephen}. This ${\it true}$ critical point $H_{c}$ may not be the
   same as the would-be first order melting point of the clean system, $H_{\text{cr}}$. It will depend on the value of the percolation threshold
   in a more microscopic calculation. 
   The experimental
   observation that $H_{c}$ is about $20$ times bigger than $H_{\text {sm}}$ is a quantitative issue and is
   beyond the scope of the present paper. A numerical simulation study of the vortex system to know
   the function $g(l)$, which involves parameters $C$ ($\sim 2\pi$ for a circle), 
    $\sigma$ and $f(p)$ is left for future study. Once maximized, this function will give
   an estimate of $H_{\text{sm}}$ ($n_{\text{sm}}=n_{\text{cr}}(p)-g(l^*)$), and will also test whether $g(l)$ is
   indeed a sharply peaked function to validate the conjecture of a sharp crossover. In Fig.~\ref{fig2}, we show a proposed phase diagram of the system within
   the randomly diluted $XY$ model.
   
   In the conventional theory of the SIT\cite{fisher1}, the Ohmic resistivity, $\rho$, was predicted to attain a universal value
   ${\it at}$ the transition point $H_c$. Its finite temperature scaling-form  was predicted to be
   \begin{equation} 
   \rho \sim \rho_Q F[\frac{|H-H_c|}{T^{1/z\nu}}],
   \end{equation}
   where
   $\rho_Q$ is the quantum of resistance, $\nu$ and $z$ are the correlation length exponent and the dynamic exponent,
   respectively, and $F$ is a scaling function. A small non-zero temperature reduces the effective dimension of the system from three
   to two, and a finite-size scaling argument
   produces the scaling-form close to the zero-temperature critical point\cite{sondhi}. Recent experiments have not found a universal critical resistivity, although
   the scaling form has been verified for $T\gtrsim 100mK$ with a value of $z\nu\sim4/3$\cite{yazdani,mason1,mason2}. At lower temperatures, breakdown
   of the scaling has been observed\cite{mason1}.

   In our model of the SIT, because $P$ is a continuous
   critical point where correlation length diverges, scaling of observables at $T=0$ is immediate and follows from the well-studied theory 
   of percolation\cite{stinch}. Scaling
   at finite temperature close to the percolation threshold is, however, not well-understood. This is because the zero-temperature
   transition is essentially geometric, and the concepts of reduced dimensionality with temperature and associated finite-size scaling don't apply. 
   Here, we will speculate on an explaination of the experimental observations. For any $T>0$ at $x=x_c$, $\rho$ is nonzero 
   since the superconducting correlation length is finite and is
   expected to be thermally activated.
   Close to the point $P$ there are two relevant correlation lengths that diverge; the percolation correlation length $\xi_{x}$ diverges
   as $\xi_{x}\sim|x-x_c|^{-\nu_x}$, and the thermal correlation length of the diluted $XY$ model, $\xi_{T}$, diverges as $\xi_{T}\sim T^{-\nu_T}$\cite{stinch}.
   This is true as long as $\xi_{x}\gtrsim \xi_{T}$, because of the reduced dimensionality of the percolating cluster. As this is a crossover
   effect for any $x\rightarrow x_c$, but not exactly ${\it at}$ the percolation threshold, the condition on $\xi_{T}$ puts a lower bound
   on temperature, only ${\it above}$ which the following holds.
   Here $\nu_x$ and $\nu_T$ are critical exponents with values in two dimensions $4/3$ and $0.98-1.03$ \cite{stinch}, respectively. In a scaled expression for 
   $\rho$ for $T \rightarrow 0$ and $x \rightarrow x_c$, these two lengths must occur in the dimensionless ratio $\xi_{T}/\xi_{x}$,
   
   \begin{equation}
   \rho(|x-x_c|, T)\sim \frac{h}{4e^2}\tilde\rho f(\xi_{T}/\xi_{x})\sim\frac{h}{4e^2}\tilde\rho \phi(\frac{|x-x_c|}{T^{\nu_T/\nu_x}}).
   \label{scaling1}
\end{equation}
   Here $\tilde\rho$ is a dimensionless constant, $\frac{h}{4e^2}$ is the quantum of resistance present for dimensional reasons,
   and $f$ and $\phi$ are scaling functions. The
   argument of $\phi$ is a power of the argument of $f$.
   Writing $H$ for $x$ and using the values for $\nu_T$ and $\nu_x$ appropriate for $2D$, we get
    
   \begin{equation}
   \rho(|H-H_c|, T)\sim \frac{h}{4e^2}\tilde\rho \phi(\frac{|H-H_c|}{T^{3/4}}).
   \label{scaling2}
   \end{equation} 
in the scaling region close to $P$.

In summary, we have proposed a low-field crossover from a strong superconductor to an inhomogenous state which is a weak superconductor and so
can be resistive even at very low temperature. If the crossover
is sharp, which is possible in two dimensions but depends on the system-parameters, it will look like a first order superconductor-metal phase
transition. The true thermodynamic superconductor-insulator critical point lies within the imhomogenous state  and is conjectured to be isomorphic to that of diluted $XY$-model.
An explaination of the observed finite temperature scaling of resistivity close to the critical point is presented and that it is only approximate, 
consistent with experiments, is pointed out.

This work has been supported by the NSF under Grant No. DMR-9971138.
I thank S. Chakravarty and S. A. Kivelson for many helpful discussions. I also thank Enzo Granato for a correspondence.


\begin{references}
\bibitem{yazdani} A. Yazdani and A. Kapitulnik, Phys. Rev. Lett. ${\bf 74}$, 3037 (1995). 
\bibitem{ephron} D. Ephron, A. Yazdani, A. Kapitulnik, and M. R. Beasley, Phys. Rev. Lett. ${\bf 76}$, 1529 (1996).
\bibitem{shimsoni} Efrat Shimshoni, Assa Auerbach, and  Aharon Kapitulnik, Phys. Rev. Lett. ${\bf 80}$, 3352 (1998).
\bibitem{mason1} N. Mason and A. Kapitulnik, Phys. Rev. Lett. ${\bf 82}$, 5341 (1999).
\bibitem{kapitulnik} A. Kapitulnik, N. Mason, S. A. Kivelson, and S. Chakravarty, Phys. Rev. B ${\bf 63}$, 125322 (2001).
\bibitem{mason2} Nadya Mason and Aharon Kapitulnik, Phys. Rev. B ${\bf 64}$, 060504(R) (2001).
\bibitem{mason3} Nadya Mason and Aharon Kapitulnik, Phys. Rev. B ${\bf 65}$, 220505(R) (2002).
\bibitem{fisher1} M. P. A. Fisher, Phys. Rev. Lett. ${\bf 65}$, 923 (1990).
\bibitem{dhlee1} D-H Lee and R. Shankar, Phys. Rev. Lett. ${\bf 65}$, 1490 (1990).
\bibitem{dhlee2} M. P. A. Fisher and D-H Lee , Phys. Rev. B ${\bf 39}$, 2756 (1989).
\bibitem{fisher2} M. P. A. Fisher, G. Grinstein, and S. M. Girvin, Phys. Rev. Lett. ${\bf 64}$, 587 (1990).
\bibitem{wen} X. G. Wen and A. Zee, Int. J. Mod. Phys. B  ${\bf 4}$, 437 (1990).
\bibitem{jaeger} H. M. Jaeger, D. B. Haviland, B. G. Orr, and A. M. Goldman, Phys. Rev. B ${\bf 40}$, 182 (1989).
\bibitem{chervenak} J. A. Chervenak and J. M. Valles, Jr., Phys. Rev. B ${\bf 59}$, 11209 (1999).
\bibitem{denis} Denis Dalidovich and Philip Phillips, Phys. Rev. Lett. ${\bf 84}$, 737 (2000); 
                D. Das and S. Doniach, Phys. Rev. B ${\bf 60}$, 1261 (1999).
\bibitem{larkin} M. Feigel'man and A. Larkin, Chem. Phys. ${\bf 235}$, 107 (1998).
\bibitem{spivak} B. Spivak, A. Zyuzin, and M. Hruska, Phys. Rev. B ${\bf 64}$, 132502 (2001).		
\bibitem{ma} Y. Imry and S. K. Ma, Phys. Rev. Lett. ${\bf 35}$, 1399 (1975).
\bibitem{imry} Yoseph Imry and Michael Wortis, Phys. Rev. B ${\bf 19}$, 3580 (1979).
\bibitem{wehr} Michael Aizenman and Jan Wehr, Phys. Rev. Lett. ${\bf 62}$, 2503 (1989).
\bibitem{dfisher} Chen Zeng, P. L. Leath, and Daniel S. Fisher, Phys. Rev. Lett. ${\bf 82}$, 1935 (1999).
\bibitem{jeffreys} H. Jeffreys, ${\textit Theory \hspace{1.0 mm}of \hspace{1.0 mm} Probability}$ (Oxford, London, 1967).
\bibitem{harris} A. B. Harris, J. Phys. C ${\bf 7}$, 1671 (1974).
\bibitem{granato} Enzo Granato and Daniel Dominguez, Phys. Rev. B ${\bf 56}$, 14671 (1997), and references therein.
\bibitem{lub} A. B. Harris and T. C. Lubensky, J. Phys. A ${\bf 17}$, L609 (1984).
\bibitem{stephen} M. J. Stephen and G. S. Grest, Phys. Rev. Lett. ${\bf 38}$, 567 (1977).
\bibitem{sondhi}S. L. Sondhi, S. M. Girvin, J. P. Carini, and D. Shahar, Rev. Mod. Phys. ${\bf 69}$, 315 (1997).
\bibitem{stinch} R. B. Stinchcombe, in {\it Phase \hspace{0.8 mm}Transitions \hspace{0.8 mm} and \hspace{0.8 mm} 
Critical\hspace{0.8 mm} Phenomena}, edited by C. Domb and J. L. Lebowitz (Academic, London, 1983), Vol. 7; D. Stauffer and A. Aharony,
{\it Introduction \hspace{0.8 mm} to \hspace{0.8 mm} Percolation \hspace{0.8 mm} Theory} (Taylor and Francis, London, 1992).		   
\end{references}
\end{document}